\newcommand{\rem}[1]{}
\documentclass[prd,twocolumn,showpacs,preprintnumbers,amsmath,amssymb]{revtex4-2}
\usepackage{amsfonts,amssymb,amsmath,mathrsfs}
\usepackage{url}
\usepackage{graphicx}
\newtheorem{thrm}{Theorem}
\newtheorem{lem}[thrm]{Lemma}
\newtheorem{prop}[thrm]{Proposition}

\newtheorem{example}[thrm]{Example}
\newtheorem{remark}[thrm]{Remark}
\begin{document}
\title[]{Conformal Killing gravity\\ 
in static spherically-symmetric spacetimes} 

\author{Carlo Alberto Mantica}
\author{Luca Guido Molinari} 
\affiliation{Physics Department Aldo Pontremoli,
Universit\`a degli Studi di Milano and I.N.F.N. sezione di Milano,
Via Celoria 16, 20133 Milano, Italy.}
\email{carlo.mantica@mi.infn.it, luca.molinari@unimi.it}

\begin{abstract} 
We identify an anisotropic divergence-free conformal Killing
tensor $K_{jl}$ for static spherically symmetric spacetimes, and write the conformal Killing gravity equations
as Einstein equations augmented by this tensor. The field equations
are of second order: this fact allows for analytic solutions and considerably simplifies the derivation of results
of previous studies based on the original Harada equations. In particular, we prove the equivalence of the known third order field equations, with the second order ones obtained by us in the conformal Killing parametrization.\\
The structure of the Ricci tensor and of the conformal Killing tensor are compatible with both anisotropic fluid sources and (non)-linear electrodynamics.
We reobtain covariantly and in simple steps the general static spherical solutions for vacuum and linear electrodynamics.
Moreover we recover the purely magnetic Lagrangian functions that induce metrics of interest for black holes.
\end{abstract}
\date{23 Jul 2024}
\maketitle

\section{Conformal Killing gravity}

Recently J. Harada \cite{Harada23,Harada23b} introduced a new theory of gravity, with field equations
\begin{align}
H_{jkl}=&8\pi\,T_{jkl} \label{eq:Harada first}\\
H_{jkl}=&\nabla_jR_{kl}+\nabla_kR_{lj}+\nabla_lR_{jk} \nonumber\\
&-\tfrac{1}{3}(g_{kl}\nabla_jR+g_{lj}\nabla_kR+g_{jk}\nabla_lR)\nonumber \\
T_{jkl}=&\nabla_jT_{kl}+\nabla_kT_{lj}+\nabla_lT_{jk} \nonumber\\
&-\tfrac{1}{6}(g_{kl}\nabla_jT+g_{lj}\nabla_kT+g_{jk}\nabla_lT)\nonumber
\end{align}
$R_{jk}$ is the Ricci tensor with trace $R$, $T_{kl}$ is
the stress-energy tensor with trace $T$. The Bianchi identity $\nabla_j R^j{}_k=\frac{1}2\nabla_k R$
implies $\nabla_j T^j{}_k =0$. Solutions of the Einstein equations
are solutions of the new theory. 

Shortly after, we found a
parametrization showing that Harada's equations are equivalent to
the Einstein equations modified by a supplemental conformal Killing 
tensor (CKT) that is also divergence-free \cite{Mantica 23 a-1}: 
\begin{align}
 & R_{kl}-\tfrac{1}2Rg_{kl}=T_{kl}+K_{kl} \label{eq:einstein enlarged}\\
 & \nabla_jK_{kl}+\nabla_kK_{jl}+\nabla_lK_{jk} \label{eq:Conformal Killing mantica-1}\\
 &=\tfrac{1}{6}(g_{kl}\nabla_jK+g_{jl}\nabla_kK+g_{jk}\nabla_lK)\nonumber
\end{align}
For this reason the theory was named Conformal Killing Gravity (CKG).
The reformulation makes Harada's extension of GR explicit through the
conformal Killing term, that satisfies $\nabla^k K_{kl}=0$ and enters as
a new source term in the equations.

Feng and Chen \cite{Feng24} proposed an action principle for CKG.
Some references on geometrical and physical applications of conformal Killing tensors are \cite{Coll06,Kobialko22,Rani 03,Sharma10}.

As $R_{kl}$ contains second
order derivatives of the metric tensor, higher orders in the field equations (\ref{eq:einstein enlarged})
may arise with the tensor $K_{jk}$. This does not occur in the present work, as well as in 
\cite{Mantica 23 a-1}, where we obtained a realization of the
conformal Killing parametrization in FRW background. 
In \cite{Mantica 23 a-1} the CKT has the perfect fluid form, candidate 
for representing the dark sector, and contains the scale factor with no derivatives. 
The Friedmann equations are thus second order in the metric, and reproduced the same forecasting 
obtained by Harada \cite{Harada23b} with eqs.\eqref{eq:Harada first}.
Vacuum cosmological
solutions, wormhole and black hole solutions were obtained by Cl\'ement and Nouicer \cite{Clement24}.
In FRW background, CKG is embraced by a general parametrization of Codazzi tensors \cite{Mantica23c}.\\
In the next work \cite{Mantica 24} we deepened the geometrical
aspects and the cosmological consequences. In particular we showed that
the density contrast in the matter era behaved as in $\Lambda$CDM and provided a fit
of the Hubble parameter versus redshift with CC and BAO, 
with forecast of future singularities. \\
Most papers on CKG dealt with a static background.
Barnes \cite{Barnes23a} found the general spherically symmetric
vacuum solution and \cite{Barnes23b,Barnes24-CqG} the general solution with
Maxwell source. Junior, Lobo and Rodriguez \cite{Junior24} investigated
regular black holes solutions of CKG coupled to nonlinear electrodynamics
and scalar fields. Further in \cite{junior 24 b} they explored black
bounce solutions in CKG coupled to nonlinear electrodynamics and
scalar fields. 
pp-wave solutions were studied by Barnes \cite{Barnes24}.\\


In this paper we show a realization of the conformal Killing
parametrization in a static spherically symmetric background
$$ ds^2 = -b^2(r) dt^2 + f_1^2(r) dr^2 + f_2^2(r) d\Omega_2^2$$
There are clear advantages in the parametrization approach \eqref{eq:einstein enlarged}: 
the results based on the equations \eqref{eq:Harada first} are here
reobtained with simplicity and covariantly.

In Section \ref{sec:Spherically-symmetric-static} we recollect the
covariant description of the Ricci tensor of a spherically symmetric
static space-time found in \cite{Mantica 23 b}. Then we show that
an anisotropic conformal Killing tensor is naturally hosted in such
spaces. It extends the perfect fluid form recently shown by Barnes in \cite{Barnes24}
in spherical symmetry.

In Section \ref{sec:CK eq  anisotropic fluid} we write the field
equations of CKG for an anisotropic fluid source: they are second
order differential equations in the metric functions, as in 
GR.\\ 
An interesting result for the metric coefficients is
obtained: if $f_2=r$ then $p_r+\mu=0$ if and only if $bf_1= (\kappa_3 r^2 + \kappa_4)^{-1/2}$.
It extends the result obtained by Barnes \cite{Barnes23b} and Cl\'ement and Noucier \cite{Clement24}
for vacuum spacetimes and linear electrodynamics.

In section \ref{sec:vacuum sol} the vacuum case is analyzed. We prove the equivalence of the third 
order differential equation \eqref{eq:Barnes third order} for $b^2(r)$ by Barnes with the second order one obtained by us, eq.\eqref{eq:general second order vacuum}.
The first one descends from Harada's equations \eqref{eq:Harada first}, the latter descends from \eqref{eq:einstein enlarged}.\\
We show
that the (unique) Schwarzschild-like solution is the vacuum by Harada 
and remark that it cannot originate from a perfect fluid CKT.

In section \ref{sec:(non)-Linear-electrodynamics} we write the CKG
equations for nonlinear electrodynamics. The same result for the metric functions in Sect. \ref{sec:CK eq  anisotropic fluid} is
proven in the nonlinear case.\\ 
In the linear case (electric and magnetic monopoles)
the general solution by Barnes \cite{Barnes23b} and Cl\'ement and Noucier \cite{Clement24} is here obtained
as solution of a second order equation.\\
For the purely magnetic case, we write the second order equation \eqref{EQ_b_MAGNETIC} for  
$b^2$ with generic magnetic Lagrangian. It straightforwardly reproduces the Lagrangians
obtained for black-hole solutions by Junior et al. \cite{Junior24}, and a generalization of the Hayward metric.

\section{Anisotropic CKT}\label{sec:Spherically-symmetric-static}
%
A covariant characterization \cite{Stephani,Mantica 22,Mantica 23 b} of static spacetimes is the existence of a time-like unit vector field that is shear, expansion, vorticity
free, with closed acceleration $\dot u_j = u^k\nabla_k u_j$:
\begin{align}
\nabla_k u_j = -u_k \dot u_j, \qquad \nabla_j \dot u_k =\nabla_k \dot u_j \label{staticeq}
\end{align}
With $\eta=\dot u^k \dot u_k$ the normalized acceleration vector is
\begin{align}
\chi_k = \frac{\dot u_k}{\sqrt\eta }
\end{align}
We focus on static spherically-symmetric spacetimes
\begin{equation}
ds^2 =-b^2(r)dt^2+f_{1}^2(r)dr^2+f_2^2(r)(d\theta^2+\sin^2\theta d\phi^2)\label{eq:Static sph symm}
\end{equation}
In this (comoving) frame: $u_0=-b$, $u_\mu=0$, $\dot u_0=0$, $\dot u_r=b'/b$, $\dot u_\theta=\dot u_\phi=0$.
A prime denotes a derivative in the variable $r$. Then: 
\begin{align}
\sqrt \eta= \frac{b'}{f_1 b}  \label{sqrteta}
\end{align}
and  $\chi_0=0$, $\chi_r=f_1$, $\chi_\theta=\chi_\phi=0$\\
The covariant expression of the Ricci tensor was obtained in \cite{Mantica 23 b}, eq.85:
\begin{align}
R_{kl}=\frac{R+4\nabla_{p}\dot u^{p}}{3}u_ku_l+\frac{R+\nabla_{p}\dot u^{p}}{3}g_{kl} \label{eq:Ricci static}\\
+\Sigma(r)\left[\chi_k \chi_l -\frac{u_ku_l+g_{kl}}{3}\right] \nonumber
\end{align}
where the coefficients are (eqs. 87, 89 of  \cite{Mantica 23 b}):
\begin{align}
  \nabla_{p}\dot u^{p}=&\frac{1}{b f_1^2}\left[b''-b'\left(\frac{f_1'}{f_1}-2\frac{f_2'}{f_2}\right )\right], \label{eq:diva acc}\\
  \Sigma(r)=&-\frac{1}{bf_1^2}\left[b''- b'\left( \frac{f_1'}{f_1}+\frac{f_2'}{f_2}\right ) \right] \label{eq:anisotropi factor}\\
 &-\frac{1}{f_1^2}\left[\frac{f_1^2}{f_2^2}+\frac{f_2''}{f_2}-\left(\frac{f_2'}{f_2}\right)^2-\frac{f_1'f_2'}{f_1f_2}\right],\nonumber
 \end{align}
$R=R^k{}_k$ and the curvature $R^\star$ of the space-like submanifold are (eqs. 90, 91 of  \cite{Mantica 23 b}): 
\begin{align}
 & R=R^\star-2\nabla_{p}\dot u^p,\\
 & R^{\star}=\frac{2}{f_2^2}-\frac{2}{f_1^2}\left[ 2\frac{f_2''}{f_2} -2 \frac{f_1'}{f_1}\frac{f_2'}{f_2}
 +\left(\frac{f_2'}{f_2}\right)^2\right ]. \label{eq:Scalar curvatures b}
\end{align}

In the background (\ref{eq:Static sph symm}) we consider
a symmetric tensor with the anisotropic structure of the Ricci tensor:
\begin{equation}
K_{kl}=\mathsf{A}(r)u_ku_l+\mathsf{B}(r)g_{kl}+\mathsf{C}(r) \chi_k\chi_l\label{eq:anisotropic Conformal killing}
\end{equation}
%
The following result shows that a static
spherically-symmetric spacetime always hosts an anisotropic
conformal Killing tensor (the proof is in Appendix 2):
\begin{thrm}\label{prop 5 Anisotropic}
In the static spherically symmetric spacetime
 (\ref{eq:Static sph symm}) the tensor (\ref{eq:anisotropic Conformal killing})
is a divergence-free conformal Killing tensor if and only if 
\begin{align*}
&\mathsf{A} = \kappa_2 f_2^2 - 2\kappa_3 b^2 , \\
&\mathsf{B} = \kappa_1+2 \kappa_2 f_2^2 + \kappa_3  b^2\\
&\mathsf{C}=-\kappa_2  f_2^2  
\end{align*}
where $\kappa_1$, $\kappa_2$ and $\kappa_3$ are arbitrary constants. 
\end{thrm}
%
$\mathsf A$, $\mathsf B$ and $\mathsf C$ are combinations 
of the metric functions $b$, $f_1$, $f_2$ with no derivatives.
Therefore, the CKG equations (\ref{eq:einstein enlarged}) and (\ref{eq:Conformal Killing mantica-1})
contain derivatives of the metric not higher than second order. 

\begin{example}[Perfect fluid CKT]
$Z_j=b u_j$ is a Kil\-ling vector, $\nabla_j Z_k + \nabla_k Z_j = 0.$
Following the strategy proposed in \cite{Rani 03}, $K_{jk}=\alpha Z_jZ_k+\beta\, g_{jk}$
with constant $\alpha$ and scalar function $\beta$, is a CKT. The divergence-free condition is $\beta=-\frac{1}{2}\alpha b^2+\gamma$. 
\begin{align*}
K_{jk}=\alpha b^2 (u_j u_k -\tfrac{1}2g_{jk}) +\gamma g_{jk} \label{eq:perfect fluid CKT}
\end{align*}
is the divergence-free CKT recently shown by Barnes in \cite{Barnes24}. It is a particular case of Theorem 1 
with $\kappa_2 =0$.
\end{example}

\section{CKG for the anisotropic fluid}\label{sec:CK eq  anisotropic fluid} 

The structures of the Ricci tensor (\ref{eq:Ricci static}) and of the CKT (\ref{eq:anisotropic Conformal killing})
 are compatible with the stress-energy tensor of an anisotropic fluid without heat flow:
\begin{equation}
T_{kl}=(\mu+p_{\perp})u_ku_l+p_{\perp}g_{kl}+(p_{r}-p_{\perp})\chi_k\chi_l
\end{equation}
$\mu$ is the energy density, $p_{r}$ and $p_{\perp}$ are the radial and transversal pressures in the comoving frame,
$\chi_k=\dot u_k/\sqrt{\eta}$. 
The tensor is rewritten as
\begin{equation*}
T_{kl}=(\mu+P)u_ku_l+Pg_{kl}+(p_{r}-p_{\perp})\left[\chi_k\chi_l-\frac{g_{kl}+u_ku_l}{3}\right]
\end{equation*}
where $P=\frac{1}{3}(p_r+2p_\perp )$ is the total pressure. The CKG equations with the fluid source are
\begin{align*}
&\frac{R+4\nabla_{p}\dot u^{p}}{3} u_ku_l+\frac{2\nabla_{p}\dot u^{p}-R}{6}g_{kl}
+\Sigma \left[\chi_k\chi_l-\dfrac{u_ku_l+g_{kl}}{3}\right]\\
&=(\mu+P)u_ku_l+ Pg_{kl}+ (p_{r}-p_{\perp})\left[\chi_k \chi_l -\dfrac{u_ku_l+g_{kl}}{3}\right]\\
&+\left[\mathsf{A}+\dfrac{\mathsf{C}}{3}\right]u_ku_l+\left[\mathsf{B}+\dfrac{\mathsf{C}}{3}\right]g_{kl}+\mathsf{C}\left[\chi_k\chi_l-\dfrac{u_ku_l+g_{kl}}{3}\right]
\end{align*}
They give three scalar equations:
\begin{align*}
&\frac{1}{3}(R+4\nabla_{p}\dot u^{p})=(\mu+P)+\mathsf{A}+\frac{1}{3} \mathsf{C}\\
&\frac{1}{6}(2\nabla_{p}\dot u^{p}-R)= P+\mathsf{B}+\frac{1}{3} \mathsf{C}\\
&\Sigma=(p_{r}-p_{\perp})+\mathsf{C}
\end{align*}
Rearranging terms and using $R=R^{\star}-2\nabla_{p}\dot u^{p}$ and Theorem 1 we obtain
\begin{prop}
The field equations of CKG in the static spherically symmetric metric (\ref{eq:Static sph symm})
with the anisotropic CKT eq.(\ref{eq:anisotropic Conformal killing}) are:
\begin{align}
\frac{R^\star}{2}&=\mu - 3\kappa_3 b^2 -\kappa_2 f_2^2 -\kappa_1 \nonumber \\ 
\nabla_{p}\dot u^{p}&= \frac{3}{2}P+\frac{1}{2}\mu + 2\kappa_2 f_2^2 +\kappa_1 \label{GCKE b} \\
\Sigma&= (p_{r}-p_{\perp})+\mathsf{C} \nonumber 
\end{align}
\end{prop}
%
%
\begin{remark}
\label{rem:Degre of freedoom} Let $f_2=r$ and consider eqs.\eqref{GCKE b} for
a perfect fluid ($p_r=p_\perp$). In GR they are $\mu =\frac{1}{2} R^\star$, $P=\frac{2}{3}\nabla_p \dot u^p - \frac{1}{6} R^\star$ and $\Sigma = 0 $: three equations for the unknowns $\mu $, $p$, $b$ and $f_1$. A further condition, such as
an equation of state, is needed. The same occurs for the conformal Killing equations. 
\end{remark}

\subsection{Properties of metric functions}
Hereafter we consider the static spherical metric with $f_2=r$, and set $f_1=h/b$. 
\begin{align}
ds^2 = - b^2(r)dt^2 + \frac{h^2(r)}{b^2(r)} dr^2 + r^2 d\Omega^2 \label{METRIC}
\end{align}
We prove a remarkable property of the metric function $h(r)$.
First we assert the following geometric result.
\begin{lem}\label{LEMMA5}
\begin{equation}
\frac{R^\star}2+\nabla_{p}\dot u^{p}+\Sigma=\dfrac{3b^2}{r}\dfrac{h'}{h^{3}}\label{eq:geometric result}
\end{equation}
\end{lem}

{\rm Proof.}
Using \eqref{eq:diva acc}, \eqref{eq:anisotropi factor}
and (\ref{eq:Scalar curvatures b}) the following relations
are straigthforwardly obtained:
\begin{align}
&R^{\star}=\dfrac{2}{r^2}+\dfrac{4b^2}{r}\dfrac{h'}{h^{3}}-\dfrac{4bb'}{rh^2}-\dfrac2{r^2}\dfrac{b^2}{h^2}\label{RST}\\
&\nabla_{p}\dot u^{p}=\dfrac{1}{h^2}(b'^2+bb'')-\dfrac{bb'}{h^2}\dfrac{h'}{h}+\dfrac{2bb'}{rh^2}\label{NABUP}\\
&\Sigma=-\dfrac{1}{h^2}(b'^2+bb'')+bb' \dfrac{h'}{h^3}+\dfrac{1}{r^2}\dfrac{b^2}{h^2}-\dfrac{1}{r^2}+\dfrac{b^2}{r}\dfrac{h'}{h^{3}}\label{eq:Scalar invariants with h}
\end{align}
The result (\ref{eq:geometric result}) follows.
\hfill $\square$

We are ready to prove  
\begin{prop}\label{prop:gen barnes} 
Consider CKG
in the metric \eqref{METRIC} with anisotropic fluid
source. Then:
\begin{equation}
\frac{1}{h^2}=\kappa_3 r^2+\kappa_4\quad\Longleftrightarrow\quad p_{r}=-\mu\label{eq:generalized barnes}
\end{equation}

{\rm Proof.}
Addition of eqs. \eqref{GCKE b} with $\mathsf C=-\kappa_2 r^2$ and $P=\frac{1}{3}(p_r+2p_\perp)$ gives
\begin{equation*}
\frac{R^\star}{2}+\nabla_p \dot u^p+\Sigma=\frac{3}{2} (p_r+\mu) - 3 \kappa_3 b^2 \label{eq:A+C for fluids}
\end{equation*}
With (\ref{eq:geometric result}) we obtain
\[
\frac{2b^2}{r}\frac{h'}{h^3} + 2\kappa_3 b^2 = p_r+\mu
\]
If $p_r + \mu=0$ it is $\dfrac{h'}{h^{3}}=-\kappa_3 r$
that integrates to (\ref{eq:generalized barnes}) with
$\kappa_4$ a constant. Conversely if $h^{-2}=\kappa_3 r^2+\kappa_4$
then $p_r+\mu=0$.
\hfill $\square$
\end{prop}
If $p_r+\mu=0$  it is
\begin{equation}
f_1 (r)=\dfrac{1}{b(r) \sqrt{\kappa_3 r^2+\kappa_4}}\label{eq: barnes}
\end{equation}
Eq.(\ref{eq: barnes}) was obtained by Barnes 
\cite{Barnes23a} while investigating vacuum solutions and also in solving CKG
field equations of linear electrodynamics \cite{Barnes23b}. 

\section{Vacuum solutions}\label{sec:vacuum sol}
In absence of matter eqs.\eqref{GCKE b}  become
\begin{align}
R^{\star}&= - 6 \kappa_3 b^2 -2\kappa_2 r^2 -2\kappa_1 \nonumber \\
\nabla_{p}\dot u^{p} &= 2\kappa_2 r^2 + \kappa_1 \label{eq:General conformal Killing vacuum}\\
\Sigma &=\mathsf{C} =-\kappa_2 r^2 \nonumber
\end{align}
It is also $h^{-2}=\kappa_3 r^2 + \kappa_4$. The remaining metric function $b^2$ is determined by $\Sigma =\mathsf C$.
With the substitution $y=b^2$ the anisotropic term (\ref{eq:Scalar invariants with h}) now is 
\begin{align}
\Sigma 
=- (\kappa_3 r^2+\kappa_4 )\frac{y''}{2}-\kappa_3 r\frac{y'}{2}+\kappa_4 \frac{y}{r^2}-\frac{1}{r^2}
\label{eq: sigma with barnes 2}
\end{align}
and the equation $\Sigma = -\kappa_2 r^2$ for $y(r)$ is:
\begin{equation}
(\kappa_3 r^2+\kappa_4) y''+ \kappa_3 ry'-2 \frac{\kappa_4}{r^2} y +\frac{2}{r^2}=2\kappa_2 r^2
\label{eq:general second order vacuum}
\end{equation}

\begin{prop} The second order equation \eqref{eq:general second order vacuum} is equivalent to the third-order equation obtained by Barnes in \cite{Barnes23a} eq.(16), with computer algebra:
\begin{align}
(\kappa_3 r^2+\kappa_4) &r^{3}y'''+ (\kappa_3 r^2 -2\kappa_4)r^2 y'' \nonumber\\
&-(\kappa_3 r^2+2\kappa_4 ) r y'+8\kappa_4 y = 8.  \label{eq:Barnes third order}
\end{align}

{\rm Proof}. The equation $\Sigma=\mathsf{C}$ gives $\Sigma'=\mathsf{C}'$.
Being $\mathsf{C}=-\kappa_2 r^2$ it is $\mathsf{C}'=\frac{2}{r}\mathsf{C}$. 
Thus the anisotropic term satisfies $\Sigma'=\frac{2}{r}\Sigma$, and 
the integral is $\Sigma=\mathsf{C}$:
\begin{equation}
\Sigma'-\dfrac2{r}\Sigma=0\quad\Longleftrightarrow\quad\Sigma=\mathsf{C}\label{eq:equivalence vacuum}
\end{equation}
The left hand side of (\ref{eq:equivalence vacuum}) is evaluated 
\begin{align*}
\Sigma'-\dfrac2{r}\Sigma
=-\dfrac{y'''}2 (\kappa_3 r^2 + \kappa_4 ) -y'' \left(\dfrac{\kappa_3 }2 r -\dfrac{\kappa_4}{r}\right ) \nonumber\\
+y'\left(\dfrac{\kappa_3}2+\dfrac{\kappa_4}{r^2}\right)-y\dfrac{4\kappa_4}{r^{3}}+\dfrac{4}{r^{3}}\label{eq:diff for sigma}
\end{align*}
and entails the equation by Barnes. \hfill $\square $
\end{prop}

The second order equation \eqref{eq:general second order vacuum} is solved ($x=r \sqrt \kappa_3$, $\kappa_4=1$):
the sum of the homogeneous (with coefficients $c_{1,2}$) and the inhomogeneous solutions
\begin{align}
b^2(r) =& c_1 \frac{\sqrt{1+x^2}}{x} + c_2 \left[\frac{\sqrt{1+x^2} }{x}{\rm arcsh}{x}  -1\right ]\\
&+1- \frac{3\kappa_2}{2\kappa_3^2}\left[\frac{\sqrt{1+x^2}}{x} {\rm arcsh}{x}-1- \frac{x^2}{3}  \right ]\nonumber
\end{align}

\begin{example}[Case $\mathbf{h=1}$] 
The vacuum static spherical solution of CKG with $h=1$ (i.e. $\kappa_3=0$ and $\kappa_4=1$) was found 
by Harada \cite{Harada23}, and extends the Schwarzschild-de Sitter solution of GR:
\begin{equation}
b^2=1-\dfrac{2M}{r}-\dfrac{\Lambda}{3}r^2-\dfrac{\lambda}{5}r^{4}\label{eq:Harada vacuum}
\end{equation}
The following result holds:
\begin{prop}
Eq.\eqref{eq:Harada vacuum} is the unique solution with $h=1$ of the vacuum equations 
\eqref{eq:General conformal Killing vacuum} with $\kappa_1=-\Lambda$, $\kappa_2=-\lambda$.

{\rm Proof.}
With $\kappa_3=0$ and $\kappa_4=1$ in \eqref{eq:general second order vacuum}, the equation $\Sigma =
\mathsf C$ is the Euler equation  $r^2y''-2y=-2+ 2\kappa_2 r^{4}$ with solution 
\begin{equation}
y=b^2=1+\dfrac{c_{1}}{r}+c_2r^2+\dfrac{1}{5}\kappa_2 r^4
\end{equation}
Thus $c_{1}=-2M$, $c_2=-\frac{1}{3}\Lambda $ and $\lambda=-\kappa_2$.\\ 
The solution must also solve the other equations in (\ref{eq:General conformal Killing vacuum}):
$R^\star =   -2\kappa_2 r^2 - 2\kappa_1$ and $\nabla_p \dot u^p = 2\kappa_2 r^2 +\kappa_1$.

With $h=1$ and $b^2=y$ eqs.\eqref{RST} and \eqref{NABUP} become: 
\begin{align*}
&R^{\star}=-\frac{2}{r}y'-\frac{2}{r^2}y  +\frac{2}{r^2} = 2\lambda r^2 + 2\Lambda\\
&\nabla_p\dot u^p=\frac{1}{2}y''+\frac{1}{r}y'=  -2\lambda r^2 -\Lambda
\end{align*}
The equations are satisfied with $\kappa_1 = -\Lambda $. \hfill $\square$
\end{prop}
The Schwarzschild de-Sitter metric occurs if and only if $\kappa_2=0$, i.e.
$\mathsf{C}=0$, giving the perfect fluid CKT. Then, 
a perfect fluid CKT in the field equations cannot originate Harada's vacuum solution \eqref{eq:Harada vacuum}.
\end{example}

\section{(non)Linear electrodynamics\\ in CKG}\label{sec:(non)-Linear-electrodynamics}

In \cite{Mantica 23 b} we specified the covariant form of the stress-energy
tensor of nonlinear electrodynamics in static spherically-symmetric
spacetimes:
%
\begin{align}
T_{jk}^{nlin}= 2 (\mathbb{E}^2+\mathbb{B}^2) \left[u_j u_k- \chi_j \chi_k \right] \mathscr L_{F}(F)
\label{eq:Stress energy non linear}\\
+2g_{jk}[\mathbb{B}^2\mathscr L_{F}(F)-\mathscr L (F)].\nonumber
\end{align}
$F=\tfrac{1}{4}F_{jk}F^{jk}$ is the Faraday scalar, $\mathscr L_{F}=d\mathscr L/dF.$
In the static setting it is $F=\tfrac{1}2(\mathbb{B}^2-\mathbb{E}^2)$, where 
\begin{prop}
\begin{align}
\mathbb{E}(r)=\dfrac{q_{e}}{r^2\mathscr L_{F}(F)}, \quad  \mathbb{B}=\frac{q_m}{r^2}
\end{align}
{\rm Proof.}
Eq.28 in \cite{Mantica 23 b} is: $\nabla_j [\chi^j \mathbb E \mathscr L_F]=\sqrt \eta \mathbb E \mathscr L_F$. Explicitly,
using \eqref{sqrteta} and the formulas in Appendix 1
$$ \left(\frac{2}{r f_1} + \frac{b'}{f_1b}\right) (\mathbb E \mathscr L_F)  + \frac{1}{f_1} \frac{d}{dr} (\mathbb E \mathscr L_F) =
\frac{b'}{f_1b}(\mathbb E \mathscr L_F) $$
Terms cancel and the linear equation gives the first result, where $q_e$ is an integration constant.\\
Eq.18 in \cite{Mantica 23 b} is $\sqrt \eta \chi^p\nabla_p \mathbb B = \mathbb B \left [\nabla_{p}\dot u^{p}-\eta-\frac{\chi^{s}\nabla_{s}\eta}{2\sqrt\eta} \right ]$. Using \eqref{eq:projector} the equation is integrated and yields the monopole solution. 
\hfill $\square$
\end{prop}
(different deductions of the proposition are in \cite{Bronnikov17,Halilsoy15}).

The tensor structure of the Ricci and of the stress-energy tensors is matched by the conformal Killing tensor. Thus
 the field equations of CKG are:
\begin{align*}
&\frac{R+4\nabla_{p}\dot u^{p}}{3}u_ku_l+\frac{2\nabla_{p}\dot u^{p}-R }{6}g_{kl}\\
&+\Sigma (r) \left[\chi_k\chi_l - \frac{u_ku_l+g_{kl}}{3}\right]
=\frac{4}{3}(\mathbb{E}^2+\mathbb{B}^2)\mathscr L_F (F)u_k u_l\\
&+2\left[\dfrac{1}{3}(2\mathbb{B}^2-\mathbb{E}^2)\mathscr L_{F}(F)-\mathscr L(F)\right]g_{kl}\\
&-2(\mathbb{E}^2+\mathbb{B}^2)\mathscr L_{F}(F)\left[\chi_k\chi_l -\frac{u_ku_l+g_{kl}}{3}\right]\\
&+\left(\mathsf{A}+\dfrac{\mathsf{C}}{3}\right)u_ku_l+\left(\mathsf{B}+\dfrac{\mathsf{C}}{3}\right)g_{kl}+\mathsf{C}
\left[\chi_k\chi_l -\frac{u_ku_l+g_{kl}}{3}\right]
\end{align*}
By equating the coefficients one obtains three scalar equations. A rearrangement of terms and use of\\ 
$R=R^{\star}-2\nabla_{p}\dot u^{p}$ give
\begin{align}
&\frac{1}{2}R^\star =2\mathscr L_{F}(F)\mathbb{E}^2+2\mathscr L(F)+\mathsf{A}-\mathsf{B}\nonumber \\
&\nabla_p \dot u^p =2 \mathscr L_F (F) \mathbb{B}^2 - 2\mathscr L(F)
+\frac{\mathsf A}{2} + \mathsf{B}+\frac{\mathsf{C}}{2}             \label{eq:CKG  nonlinear electro}\\
&\Sigma=-2(\mathbb{E}^2+\mathbb{B}^2)\mathscr L_{F}(F)+\mathsf{C}\nonumber
\end{align}

%
Also in this case we are able to show the validity of (\ref{eq: barnes}).
The following result holds:
\begin{prop}\label{prop:gen barnes -1}
Consider CKG  coupled with nonlinear
electrodynamics in the metric (\ref{METRIC}). Then 
$$\frac{1}{h^2}=\kappa_3 r^2+ \kappa_4$$

{\rm Proof.}
The sum of eqs. (\ref{eq:CKG  nonlinear electro})  is
$\tfrac{1}{2}R^\star +\nabla_p \dot u^p + \Sigma=\tfrac{3}{2} (\mathsf{A}+\mathsf{C})$. 
Now use Lemma \ref{LEMMA5} and $\mathsf{A}+\mathsf{C}=- 2 \kappa_3 b^2$.
\hfill $\square$
\end{prop}

\subsection{Linear electrodynamics}
In linear electrodynamics $\mathscr L(F)=F=\tfrac{1}2(\mathbb{B}^2-\mathbb{E}^2)$. 
The electric field is Coulomb $\mathbb{E}(r)=q_e/r^2$, so that 
$$\mathbb{E}^2+\mathbb{B}^2=\frac{q^2}{r^{4}}$$ 
with  $q^2=q_e^2+q_m^2$. Equations (\ref{eq:CKG  nonlinear electro}) take the form
\begin{align}
&\frac{1}{2}R^\star = \frac{q^2}{r^4}-3\kappa_3 b^2 -\kappa_2 r^2 -\kappa_1 \label{RSTARLIN}\\
&\nabla_p \dot u^p =  \frac{q^2}{r^4} +2\kappa_2 r^2 + \kappa_1 \label{NABLAULIN}\\
&\Sigma=-2 \frac{q^2}{r^4}+\mathsf{C}  \label{SIGMALIN}
\end{align}
$\Sigma $ is given by (\ref{eq: sigma with barnes 2}) and $\mathsf{C}=-\kappa_2 r^2$. 
Eq.\eqref{SIGMALIN} is the second order differential equation for $y=b^2$:
\begin{equation}
(\kappa_3 r^2+\kappa_4 ) y'' +\kappa_3 r y'
-2 \kappa_4 \dfrac{y}{r^2}+\dfrac2{r^2}=2\kappa_2 r^2+\dfrac{4q^2}{r^{4}}\label{eq:secon order linear electrodyn}
\end{equation}

\begin{prop}
In CKG coupled with linear electrodynamics the second order equation for $b^2(r)$ in a static
spherically symmetric background is equivalent to the following third order equation:
\begin{align}
&(\kappa_3 r^2+\kappa_4) r^{3}y'''+ (\kappa_3 r^2-2\kappa_4 ) r^2y''  \label{eq: Third order lin electro}\\
&\qquad - (\kappa_3 r^2+2\kappa_4) ry'+8\kappa_4 y =8-\dfrac{24q^2}{r^2} \nonumber
\end{align}
This is eq.14 worked by Barnes in \cite{Barnes23b} and the ``master equation'' by Cl\'ement and Noucier \cite{Clement24}.

{\rm Proof.} As in the vacuum case, $\Sigma=-2\frac{q^2}{r^4}+\mathsf{C}$
gives $\Sigma'=\frac{8q^2}{r^5}+\mathsf{C}'$. Since
$\mathsf{C}=-\kappa_2 r^2$ it is $\mathsf{C}'=\frac{2}{r}\mathsf{C}=\frac{2}{r}\left[\Sigma+2\frac{q^2}{r^4}\right]$.
Thus the anisotropic term satisfies $\Sigma'-\frac2{r}\Sigma=\frac{12q^2}{r^{5}}$;
on the other hand the integral is $\Sigma=-2\frac{q^2}{r^{4}}+\mathsf{C}$. The following equivalence holds: 
\begin{equation}
\Sigma'-\dfrac2{r}\Sigma=\dfrac{12q^2}{r^{5}}\quad\Longleftrightarrow\quad\Sigma=-2\dfrac{q^2}{r^{4}}+\mathsf{C}\label{eq:equivalence  electrodynamics}
\end{equation}
Using \eqref{eq: sigma with barnes 2} the left hand side of (\ref{eq:equivalence  electrodynamics}) is evaluated and gives 
the third order equation. \hfill $\square$
\end{prop}

As a check, the second order equation \eqref{eq:secon order linear electrodyn} is solved for $\kappa_4=1$, 
$x=r\sqrt \kappa_3$. 
The solution, evaluated with Mathematica, coincides with that by Cl\'ement and Noucier \cite{Clement24}
\begin{align*}
y(x)&= c_1 \frac{\sqrt{1+x}}{x} + c_2 \left[ \frac{ \sqrt{1+x^2}}{x} {\rm arcsh}(x)-1\right ] +1\\
&-\frac{3\kappa_2}{2\kappa_3^2}\left [ \frac{ \sqrt{1+x^2}}{x} {\rm arcsh}(x)-1-\frac{x^2}{3}\right ]
+ \kappa_3 q^2(2+\frac{1}{x^2}) 
\end{align*}

\begin{example}[Case $\mathbf{h=1}$] 
The metric function 
\begin{equation}
b^2(r)=1-\dfrac{2M}{r}-\dfrac{\Lambda}{3}r^2+\dfrac{q^2}{r^2}-\dfrac{\lambda}{5}r^{4}\label{eq:junior barnes}
\end{equation}
was obtained in \cite{Junior24,Barnes23b}
with different strategies. In \cite{Junior24} Junior, Lobo and
Rodriguez posed a functional form of $b(r)$ including
the term $-\frac{\lambda}{5}r^{4}$ that characterizes CKG, and
then engineered a nonlinear electrodynamics Lagrangian and its derivative.
On the other hand, in \cite{Barnes23b} Barnes used computer algebra
to obtain the same and more general solutions.\\
Here the solution (\ref{eq:junior barnes}) is achieved in a much simpler way:
\begin{prop}
The unique solution of \eqref{RSTARLIN}--\eqref{SIGMALIN} with $h=1$ (i.e. $\kappa_3=0$, $\kappa_4=1$)
is the metric function (\ref{eq:junior barnes}), with $\kappa_1=-\Lambda$, $\kappa_2=-\lambda$.

{\rm Proof.}
The equation $\Sigma =- 2\frac{q^2}{r^4}+\mathsf C$ with $\Sigma $ in \eqref{eq: sigma with barnes 2}
and $y=b^2$ is Euler's equation $r^2y''-2y+2 = 4\frac{q^2}{r^2}+2\kappa_2 r^4$ with solution 
\begin{equation}
y=b^2=1+\frac{c_{1}}{r}+c_2r^2+\frac{q^2}{r^2}+\dfrac{\kappa_2}{5} r^4. 
\end{equation}
This is \eqref{eq:junior barnes} with $c_1=-2M$, $c_2=-\frac{\Lambda}{3}$ and $\lambda=-\kappa_2$.\\
Let's show that it also solves \eqref{RSTARLIN} and \eqref{NABLAULIN}:
\begin{align*}
&\tfrac{1}{2}R^\star = q^2/r^4  -\kappa_2 r^2 - \kappa_1\\
&\nabla_p \dot u^p =  q^2/r^4 +2\kappa_2 r^2 +\kappa_1
\end{align*}
With \eqref{eq:junior barnes}, $h=1$ and $b^2=y$, eqs.\eqref{RST} and \eqref{NABUP} give:
\begin{align*}
&\tfrac{1}{2}R^\star = \tfrac{1}{r^ 2} - \tfrac{1}{r} y' - \tfrac{1}{r^2}y  = q^2/r^4 -\kappa_2 r^2 -3c_2\\
&\nabla_p \dot u^p = \tfrac{1}{2} y''+\tfrac{1}{r}y' = q^2/r^4 + 2\kappa_2 r^2 + 3c_2
\end{align*}
Equality is achieved with $\kappa_1=3c_2$.
\hfill $\square$
\end{prop}
\end{example}

\begin{example}
For $\kappa_3=1$ and $\kappa_4=0$ eq.\eqref{eq:secon order linear electrodyn} is solved by
\begin{equation}
y=\lambda-\dfrac{\Lambda}{3}r^{2}+m\ln r-\dfrac{1}{2r^{2}}+\dfrac{q^{2}}{4r^{4}}
\end{equation}
This is eq. 19 in Barnes or 2.21 in Cl\'ement and Nouicer.
\end{example}

\subsection{Nonlinear electrodynamics: \\purely magnetic solutions}
In this case $\mathbb{E}(r)=0$ and $\mathbb{B}=q_m/r^2$ (magnetic
monopole). Then $F=\frac{\mathbb{B}^2}{2}=\frac{q_m^2}{2r^4}.$
The third equation of (\ref{eq:CKG  nonlinear electro}) is 
\begin{align*}
\Sigma=-2\dfrac{q_m^2}{r^4} \mathscr{L}_F (F)+\mathsf{C}
\end{align*}
Now $\mathscr{L}_{F}=\dfrac{d\mathscr{L}}{dF}=\dfrac{d\mathscr{L}}{dr}\dfrac{dr}{dF}=- \mathscr{L}' \frac{r^5}{2q_m^2}$,
so that the previous equation rewrites as 
\begin{equation}
\Sigma=r\mathscr{L}'+\mathsf{C}.   \label{eq:non linear purely magnetic}
\end{equation}
The knowledge of $\mathscr L(r)$ in (\ref{eq:non linear purely magnetic})
gives a second order differential equation for the metric function $b^2=y$:
\begin{align}
(\kappa_3r^2+\kappa_4) y'' + \kappa_3 r y' -\frac{2\kappa_4}{r^2} y +\frac{2}{r^2}
=-2(r\mathscr{L}'-\kappa_2 r^2) \label{EQ_b_MAGNETIC}
\end{align}

\begin{example}[Case $\mathbf{h=1}$] For $\kappa_3=0$ and $\kappa=1$ the solution by the method of variation of parameters
is:
\begin{align}
y(r) =1+ c_1r^2 + \frac{c_2}{r} -\frac{\kappa_2}{5} r^4 - \frac{2}{r} \int^r dr' r'^2 \mathscr{L}(r')
\end{align}
Given ${\mathscr L}(F)$, $F=q_m^2/(2r^4)$, one evaluates $b(r)$.\\ 
In \cite{Junior24} a different strategy is pursued:
an input metric with faithful properties (such as regularity in the origin) is chosen, and the corresponding Lagrangian
is reconstructed. \\
Here this procedure is greatly facilitated by eq.(\ref{eq:non linear purely magnetic}). It is illustrated to reproduce two interesting examples:\\
$\bullet $ Metric eq.34 in \cite{Junior24}:
\begin{equation}
y = 1-\frac{2M}{r} - \frac{\Lambda}{3}r^2 + \frac{q_m^2}{r^2} - \frac{\lambda}{5}r^4+\frac{k_0}{r^4}
\label{eq:metric lobo 1}
\end{equation}
With $\Sigma=\dfrac{y-1}{r^2}-\dfrac{1}{2}y''$
we get $\Sigma=\lambda r^{2}-\dfrac{2q_{m}^{2}}{r^{4}}-9\dfrac{k_{0}}{r^{6}}$.
Eq.(\ref{eq:non linear purely magnetic}) is easily solved by
\begin{align}
\mathscr{L}&=\phi_0+ \frac{q_{m}^{2}}{2r^{4}}+\phi_{1}r^{2}+3\frac{k_0}{2r^{6}} \nonumber \\ 
&= \phi_0 + F +\phi_{1}\frac{q_m}{\sqrt{2F}}+3\frac{k_0 (2F)^{3/2}}{2q_m^3}
\end{align}
with constants $\phi_0$, $\phi_1$. This is eq 41 in \cite{Junior24}.\\
$\bullet $ Bardeen-type metric eq.72 in \cite{Junior24}): 
\begin{equation}
y=1-\frac{2Mr^2}{(q_m^2+r^2)^{3/2}}-\frac{\Lambda}{3}r^2 - \frac{\lambda}{5} r^4 \label{eq:metric lobo 1-1}
\end{equation}
Now $\Sigma=\lambda r^{2}-\dfrac{15Mr^{2}}{(q_{m}^{2}+r^{2})^{7/2}}$. Eq.(\ref{eq:non linear purely magnetic}) is easily solved:
\begin{equation}
\mathscr{L}=\dfrac{3Mq_{m}^{2}}{(q_{m}^{2}+r^{2})^{5/2}}+\phi_{1}r^{2}+\phi_{0}\label{eq: lagrangian 1 lobo-1}
\end{equation}
with $\phi_0$ and $\phi_1$ constants. This is eq.73 in \cite{Junior24}. With $F=q_m^2/(2r^4)$, 
$\mathscr L$ can be rewritten as 
\begin{equation}
\mathscr{L}(F)=\frac{3Mq_m^2}{\left(q_m^2+\dfrac{q_{m}}{\sqrt{2F}}\right)^{5/2}}+\phi_{1}\dfrac{q_{m}}{\sqrt{2F}}+\phi_{0}
\end{equation}
This is eq.77 presented in \cite{Junior24}.
\end{example}

\noindent
This is a new example:\\
$\bullet$ Hayward-like solution. The metric 
\begin{equation}
b^{2}=1-\frac{2Mr^2}{q_m^3+r^3}-\frac{\Lambda}{3}r^2 -\frac{\lambda}{5}r^4\label{eqhayward like}
\end{equation}
is the Hayward black-hole when $\lambda,\Lambda=0$ \cite{Hayward96}. It is inferred that $\Sigma=\lambda r^{2}-\dfrac{18Mr^{3}q_{m}^{3}}{(q_{m}^{3}+r^{3})^{3}}.$
Then (\ref{eq:non linear purely magnetic}) is solved by
\begin{align}
\mathscr{L}=&\frac{3Mq_m^3}{(q_m^3+r^3)^2}+\phi_1r^2 +\phi_0 \nonumber \\
=&\frac{3M(2F)^{3/2}}{\left(1+(2Fq_m^2)^{3/4}\right)^2}+\phi_1 \frac{q_m}{\sqrt{2F}}+\phi_0  \label{eq: lagrangian hayward}
\end{align}

\section{Conclusions}
We have shown that static spherically symmetric spacetimes naturally host an anisotropic divergence-free conformal
Killing tensor (Theorem 1). This makes the parametrization \eqref{eq:einstein enlarged} of the Harada equations as modified Einstein equations effective for such background.\\ 
The CKG equations can support an anisotropic fluid source as well as (non)-linear electrodynamics.
In both cases the equations are second order. This is a great advantage with respect to the solution by components
of the third order Harada equations found in the existing literature. We prove the equivalence of our second order equations
with the third order ones.
Our approach recovers several results obtained so far in a simple and covariant way, and gives some new ones.

\section*{Appendix 1. Useful formulae} 
We collect useful formulae for static spherical spacetimes, taken from \cite{Mantica 23 b}.\\ 
A scalar function $F(r)$ has gradient in the radial direction, given by the unit spacelike vector $\chi$:
\begin{align*}
\nabla_j F = \chi_j \chi^k \nabla_k F = \chi_j \frac{F'}{f_1}
\end{align*}
(the prime is $d/dr$). In particular: 
\begin{align}
\frac{\chi^s\nabla_s \eta}{2\sqrt\eta} =  \frac{1}{2\sqrt{\eta}}\frac{1}{f_1}\frac{d}{dr} \left(\frac{b'^2}{b^2f_1^2} \right )  \nonumber\\
=\frac{1}{f_1^2}\left[ \frac{b''}{b} - \frac{b'^2}{b^2}  - \frac{b'f_1'}{bf_1} \right ] \label{gradeta}
\end{align}
The vector $\chi_k=\dot u_k/\sqrt{\eta}$ is normalized, $\chi^k\chi_k=1$. 
\begin{equation}
\nabla_j\chi_k=\frac{\nabla_j\dot u_k}{\sqrt{\eta}} -\frac{\chi^{s}\nabla_{s}\eta}{2\eta}\chi_j\chi_k \label{eq:grad chi}
\end{equation}
Since $\dot u_k$ is closed, also $\chi_j$ is closed: $\nabla_j\chi_k=\nabla_k\chi_j$. A consequence is $\chi^j\nabla_j\chi_k=\chi^j\nabla_k\chi_j=0$.\\
The gradient of the acceleration is (eq.16 in \cite{Mantica 22}):
\begin{align}
\nabla_j\dot u_k=& -\eta u_k u_j + \frac{\chi^s\nabla_s\eta}{2\sqrt\eta}\chi_j\chi_k \label{eq:acceleration and eta}\\
 &+ \frac{1}{2}N_{jk} \left[\nabla_{p}\dot u^{p}-\eta-\frac{\chi^s\nabla_{s}\eta}{2\sqrt \eta}\right] \nonumber
\end{align}
where $N_{jk} = g_{jk} + u_j u_k - \chi_j \chi_k$ is a projection. \\
Eq. 87 in \cite{Mantica 23 b},  eqs.\eqref{sqrteta} and \eqref{gradeta} give:
\begin{equation}
\nabla_{p}\dot u^{p}-\eta-\frac{\chi^{s}\nabla_{s}\eta}{2\sqrt\eta}=\frac2{f_{1}^2}\left[\frac{b'}{b}\frac{f_2'}{f_2}\right]\label{eq:projector}
\end{equation}
One then obtains:
\begin{align}
&\nabla_j \chi_k =  \frac{1}{f_1}\left[\frac{f_2'}{f_2}-\frac{b'}{b}\right ]  u_k u_j  + \frac{f_2'}{f_1f_2}(g_{jk} - \chi_j \chi_k) \label{nablachi}\\
&\nabla_p\chi^p =  2\frac{f_2'}{f_1f_2}+\frac{b'}{f_1 b} \label{eq:div chi def-1}
\end{align}

\section*{Appendix 2}
Let $K_{kl} = \mathsf{A} u_k u_l + \mathsf{B} g_{kl} + \mathsf{C} \chi_k \chi_l $. \\
The Conformal Killing condition \eqref{eq:Conformal Killing mantica-1} with the static equation \eqref{staticeq} for $u_i$ and the closedness of $\chi_i$ is:
\begin{align*}
0 =&\nabla_i (\mathsf{B}-\tfrac{1}{6}K) g_{jk} + ( \nabla_j\mathsf{B} -\tfrac{1}{6}K) g_{ki} 
+(\nabla_k \mathsf{B} -\tfrac{1}{6}K) g_{ij}  \\
&+(\nabla_i \mathsf{A} -2\mathsf{A} \dot u_i ) u_j u_k + (\nabla_j \mathsf{A}-2\mathsf{A} \dot u_j ) u_k u_i  \nonumber\\
&+ (\nabla_k \mathsf{A}-2\mathsf{A}\dot u_k ) u_i u_j \nonumber\\
&+ (\nabla_i \mathsf C)\chi_j \chi_k + (\nabla_j \mathsf C)\chi_k \chi_i + (\nabla_k \mathsf C)\chi_i \chi_j \nonumber\\
&+ 2\mathsf C (\chi_i \nabla_j \chi_k + \chi_j \nabla_k \chi_i +\chi_k \nabla_i \chi_j )\nonumber
\end{align*}
Since $\mathsf A$ only depends on $r$, it is $\nabla_i \mathsf{A} = \chi_i \chi^k\nabla_k \mathsf A =\chi_i {\mathsf A}'/f_1$, and similarly for $\mathsf B$ and $\mathsf C$ and $K $. The equation, multiplied by $f_1$  becomes:
\begin{align}
0=&(\mathsf{B}'-\tfrac{1}{6}K' ) (\chi_i g_{jk} + \chi_jg_{ki} +\chi_k g_{ij} )\label{MAIN} \\
+&(\mathsf{A}' -2\frac{b'}{b}\mathsf{A}) (\chi_i u_j u_k +\chi_j u_k u_i  + \chi_k u_i u_j) \nonumber\\
+&3 \mathsf {C}' \chi_i\chi_j \chi_k 
+ 2f_1 \mathsf C (\chi_i \nabla_j \chi_k + \chi_j \nabla_k \chi_i +\chi_k \nabla_i \chi_j )\nonumber
\end{align}
Contraction with $\chi^i\chi^j\chi^k$:
\begin{align}
0=3\mathsf{B}'+ 3{\mathsf C}'-\tfrac{1}{2}K'  =  \tfrac{1}{2}({\mathsf A}' + 2{\mathsf B}' + 5 {\mathsf C}') \label{ABC}
\end{align}
The identity is used to simplify $K'$ and ${\mathsf A}'$ from the equation:
\begin{align*}
0=&(\mathsf{A}' -2\frac{b'}{b}\mathsf{A}) (\chi_i u_j u_k +\chi_j u_k u_i  + \chi_k u_i u_j) \nonumber\\
+&\mathsf {C}' [3\chi_i\chi_j \chi_k - (\chi_i g_{jk} + \chi_jg_{ki} +\chi_k g_{ij})]\\
+& 2f_1 \mathsf C (\chi_i \nabla_j \chi_k + \chi_j \nabla_k \chi_i +\chi_k \nabla_i \chi_j )\nonumber
\end{align*}
Contraction with $u^i u^j \chi^k$:
$0=(\mathsf{A}' -2\frac{b'}{b}\mathsf{A})    +\mathsf {C}'  + 2f_1 \mathsf C   u^j\dot \chi_j $. It is
$ u^j\dot \chi_j = -\dot u^j \chi_j =-\sqrt \eta = - b'/(bf_1)$. Then: 
$\mathsf{A}' +\mathsf{C}' = 2\frac{b'}{b}(\mathsf{A} + \mathsf C)$, 
with solution $\mathsf{A} + \mathsf{C} = -2\kappa_3 b^2(r)$.\\
Contraction with $g^{ij}\chi^k$:
\begin{align*}
0=&-\mathsf{A}' + 2\frac{b'}{b}\mathsf{A} - 3\mathsf {C}'   + 2f_1 \mathsf C  \nabla^j \chi_j 
\end{align*}
Use \eqref{eq:div chi def-1} and obtain:
\begin{align*}
0=&-(\mathsf{A}' + \mathsf {C}' )  + 2\frac{b'}{b}(\mathsf{A}+ \mathsf C) - 2\mathsf {C}' + 4\frac{f_2'}{f_2}\mathsf C 
\end{align*}
It follows that ${\mathsf C} =-\kappa_2 f^2_2$. Then ${\mathsf A}= -2\kappa_3 b^2 +\kappa_2 f^2_2$. Eq.\eqref{ABC} gives
${\mathsf B} = \kappa_3b^2 +2 \kappa_2 f^2_2 + \kappa_1$ and $K= 6\kappa_3 b^2 +6\kappa_2 f^2_2 + 4\kappa_1$.\\
The found parameters $\mathsf{A,B,C}$ are inserted in \eqref{MAIN}. Up to a factor $2\kappa_2 f_1f_2^2$ it is: 
\begin{align}
0=&\frac{f_2'}{f_1f_2}  (\chi_i g_{jk} + \chi_jg_{ki} +\chi_k g_{ij} ) \label{GOLD}\\
+&\left [ \frac{f_2'}{f_1f_2} - \frac{b'}{bf_1}\right ] (\chi_i u_j u_k +\chi_j u_k u_i  + \chi_k u_i u_j) \nonumber\\
-&3\frac{f_2'}{f_2f_1} \chi_i\chi_j \chi_k 
-  (\chi_i \nabla_j \chi_k + \chi_j \nabla_k \chi_i +\chi_k \nabla_i \chi_j )\nonumber
\end{align}
With the expression \eqref{nablachi}
 the conformal Killing equation is satisfied for any choice of the constants.
 
Now we prove the opposite statement: in the metric \eqref{eq:Static sph symm}, if $\mathsf{A,B,C}$ are those in thrm.1, then the tensor \eqref{eq:anisotropic Conformal killing} is conformal Killing. This is expressed by condition \eqref{MAIN}, that becomes \eqref{GOLD} after substitutions. The latter equation is identically satisfied.
  \hfill $\square $

\subsection*{Data availability statement}
No data are available because of the nature of the research.

\end{document}